\documentstyle[prb,twocolumn,aps]{revtex}
\begin{document}
\draft

\title{\bf $\eta$-superconductivity in the one-dimensionnal pair
hopping model. }
\author{\bf Georges Bouzerar\cite{byline1} and G. I. Japaridze
\cite{byline2}
}
\address{ Inst. f\"ur Theoretische Physik, Univ. zu K\"oln,\\  
Z\"ulpicher Stra\ss e 77, K\"oln 50937, Germany.}

\date{May 1996}
\maketitle

\begin{abstract} 
Using exact Lanczos diagonalizations
we have shown that the pair-hopping model for {\it negative}
$W$ exhibits a phase transition
into $\eta$-{\it superconducting state}.
The transition occurs at {\it any band filling} and
the critical value $W_{c}$ varies between 
$W_{c} \simeq -1.75t$ at half-filling
to $W_{c} \simeq -2t$ for two particles on
the lattice.
\end{abstract}

   There is considerable interest in the pair--hopping model as a 
phenomenological model which captures some of the essential physics of the 
high--$T_{c}$ superconducting materials. The model was proposed ten years ago 
by Penson and Kolb \cite{PK} as a model exhibiting the real--space pair 
formation mechanism. The Hamiltonian of the pair--hopping model contains, 
in addition to the usual 
one--electron hopping term, a term that hops singlet pairs of electrons from 
site to site and in the one--dimensional case is given by
\begin{eqnarray}\label{1.1}
{\cal H} & = & -t\sum_{n,\alpha}(c^{\dagger}_{n,\alpha}c_{n+1,\alpha}
+c^{\dagger}_{n+1,\alpha}c_{n,\alpha})\nonumber\\   
& - & W\sum_{n}(c^{\dagger}_{n,\uparrow}c^{\dagger}_{n,\downarrow}
c_{n+1,\downarrow}c_{n+1,\uparrow} + h.c.).
\end{eqnarray}
Here $c^{\dagger}_{n,\alpha}$  ($c_{n,\alpha}$) is the creation (annihilation)
operator for an electron with spin ${\alpha}$ at site $n$. 
Thus $t$ and $W$ are the single--electron and pair--hopping amplitudes 
respectively, and the competition between these two sources of delocalization 
leads to a rich phase diagram of the model \cite{PK,AM,HD,SA,BB}.
For $W=0$, we have a 
system 
of free electrons, whose properties are exactly  known. In the opposite limit 
$t=0$ all sites are doubly occupied or empty, the model is superconducting by 
construction and equivalent to the $XY$ model\cite{PK}. The correspondance 
with the $XY$ model can be seen by introducing $S=1/2$ pseudospins 
$T_{n}$, with $T^{+}_{n} =  c^{\dagger}_{n,\uparrow}c^{\dagger}_{n,\downarrow},T^{z}_{n} =  (c^{\dagger}_{n,\uparrow}c_{n,\uparrow}+
c^{\dagger}_{n,\downarrow}c_{n,\downarrow}-1)/2.$
Then the Hamiltonian becomes
\begin{equation}\label{1.3}
{\cal H} =  - W\sum_{n}(T^{+}_{n}T^{-}_{n+1}+ h.c).
\end{equation}
This picture holds true for either sign of $W$, but it is important to note 
that $W \rightarrow -W$ is not a symmetry of the model (\ref{1.1}) 
and the way the system reaches its limiting 
behaviour at $\left|W\right|>>t$ is genuinely different for negative and 
positive $W$ cases \cite{AM}. Moreover, as we show in this paper, if for 
{\it positive}--$W$ a singlet superconducting state with order 
parameter $\Delta_{SS}=\langle c^{\dagger}_{k,\uparrow}
c^{\dagger}_{-k,\downarrow}\rangle$ corresponding to the usual Cooper pairs
 is realized \cite{AM}, in the case of 
{\it negative}--$W$, an $\eta$--{\it superconducting} state \cite{CNY} with 
order parameter $\Delta_{\eta}=\langle c^{\dagger}_{k,\uparrow}
c^{\dagger}_{\pi-k,\downarrow}\rangle $ corresponding to the  
pairing with total momentum equal to $\pi$ is the ground state of the 
system at $W<W_{c} \approx -1.75t$.

The model was mainly studied in the case of half--filled band and $W>0$. 
Penson and Kolb used exact diagonalizations data for 
chains up to 10 sites and found a phase transition into a superconducting 
state at which the spin gap (or single--particle excitation gap) opens for 
$W>W_{c} \approx 1.4t$. Later Affleck and Marston\cite{AM} analysed 
the model within the framework of the weakly--coupling continuum limit 
approach. They found that for any $W>0$ the spin excitation spectrum is 
gapped, while the charge excitation spectrum is gapless and the singlet 
superconducting instabilities are most divergent in the ground--state. 
In the case of $W<0$ ($\mid W \mid \ll t$)
at half--filling there is a gap in the charge excitation 
spectrum, the spin sector is gapless and the ground state corresponds to an 
insulator. Moreover Affleck and 
Marston argued the absence of any other transitions for $W>0$ and the 
necessarity 
of a phase transition into a sector with gapped spin excitations and 
gapless charge degrees of freedom for $-W \gg  t$. This scenario of the ground--state phase diagram was recently confirmed by Sikkema and Affleck \cite{SA} 
who used the Density Matrix Renormalization Group technique
(DMRG) for chains 
up to 60 sites. For $W < 0$, they found a transition into a spin gapped  
phase at $W< W_{c} \approx -1.5t$.

In this paper we focus our attention on the transition into an $\eta$-superconducting 
state for {\it negative}--$W$. This transition corresponds to a drastic 
change of the ground-state,
after which the one particle hopping term is almost frozen out.
To prove this,
we used exact Lanczos diagonalizations.
We have shown that the transition into 
$\eta$-superconducting state takes place at {\it arbitrary filling} and
{\it weakly} depends on the band--filling.
Our data suggests that the finite size effect are extremely
small, that is why we restricted ourself
to systems up to 10 sites.

As far as the main phenomenon characterizing the transition into an  
$\eta$--superconducting state takes place in the momentun space it is 
convenient to rewrite the Hamiltonian in this space
\begin{equation}\label{1.4}
{\cal H} = -2t\sum_{k,\alpha}c^{\dagger}_{k,\alpha}c_{k,\alpha}\cos(k)
 -  2W\sum_{Q}A^{\dagger}_{Q}A_{Q}\cos(Q)
\end{equation}
where $A^{\dagger}_{Q} = {1 \over \sqrt{L}}\sum_{k}c^{\dagger}_{k,\uparrow}
c^{\dagger}_{Q-k,\downarrow}$
is the creation operator of pair of electrons with 
opposite spins and total momentum $Q$, $L$ is the size of the 
system.
As far as the total momentum of the system $Q_{tot}$ is conserved 
we can consider each sector of the Hilbert space for a given $Q_{tot}$ 
independently. It is clear that the ground state of the system belongs to 
the sectors $Q_{tot}=0$ or $Q_{tot}=\pi$. States with 
$Q_{tot} \neq 0$ and $\pi$ exhibit a broken 
time reversibility symmetry and are excited states.

As it was shown by Yang \cite{CNY} the general condition
ensuring the possibility for $\eta$--pairing is
\begin{equation}\label{1.5a}
\epsilon(\vec k)+ \epsilon(\vec \pi-\vec k)={\it constant} 
\end{equation}
where $\epsilon(\vec k)$ is the bare electron spectrum
and $\vec \pi = (\pi,..,\pi)$.
The particularity of the pair-hopping
model could be observed even in the case
of two particles on a lattice.
Below we consider the one-dimensional case, however
the arguments are valid in any dimensions if
the bare electron spectrum ensures the condition (\ref{1.5a}).

Let us first analyse the $W < 0$ case. 
For $W = 0$ the ground-state energy
is $E_{0} =-4t$ and the total momentum
is $Q_{tot}= 0$.
As we switch on W, it can be shown easily
that the lowest energy in the $Q_{tot}= 0$
subspace goes to $-2W > 0$ (for $ | W | \gg t$).
On the other hand, the lowest energy in the 
$Q_{tot}= \pi$ sector is $2W < 0$.
Thus there should be a transition at some critical value,
from the unpaired state in the $Q_{tot}= 0$ sector into the 
$\eta$-pairing state in the $Q_{tot}= \pi$ sector.
We can roughly estimate the critical value
to be $W_{c} \simeq -2t$.
Indeed after the transition the ground-state
corresponds to the wave function $ A^{\dagger}_{\pi} | 0 >$,
which clearly describes an $\eta$-pair.
It is clear that after the transition the one particle hopping term
is {\it completely} frozen out (i.e $< n(k) > = \sum_{\alpha}
< c_{k \alpha}^{\dagger}c_{k \alpha} >=
constante $).

In the opposite case $ W > 0$, the ground-state
{\it always} remains in the $Q_{tot}= 0$ subspace and its
energy continuously
goes from $-4t$ at $W = 0$ to $-2W$ for $W \gg t$.
There is no more transition for $ W > 0$ and the weight
of Cooper pair {\it continuously} increases
up to 1 when $W =+\infty$.
In this limit the ground state wave function is
 $ A^{\dagger}_{0} | 0 >$. Note that only for $W=\pm \infty$ the 
ground-states 
are equivalent up to a trivial $\pi/2$ shift of all 
momenta, reflecting the 
$W \rightarrow -W$ symmetry of the $XY$ model.

As we show below using exact diagonalizations data
the picture essentially remains in 
the case of many particle systems. Namely for $W < 0$,
there is a phase transition (presumably of the first order) 
at the critical value of the pair--hopping amplitude 
from the insulating 
(1/2-filled case) or Luttinger--liquid state (lower
filling) into a $\eta$--superconducting state. 
The transition occurs at {\it any band filling} and
the critical value $W_{c}$ varies between 
$W_{c} \simeq -1.75t$ at half-filling
to $W_{c} \simeq -2t$ for two particles on
the lattice.
The transition to $\eta$-pairing is characterized
by the strong reduction
of the one particle hopping term.
However, it should be stressed that due to
 the quasi-bosonic character of the pairs,
the weight of unpaired particles is extremely small but {\it finite}
after the transition.

In the following part, we will present numerical data
considering the lowest energy in the $Q=0$ and $Q=\pi$
sectors.
In Fig.1 we have plotted the lowest energy in each sectors 
at half-filling, as a function of W
for two different cases $L=8$ and $L=10$.
In Fig.1a we clearly observe a different behaviour of the ground-state
energy for $W >0$ and $W < 0$.
In the case $W > 0$ the ground state energy continuously
goes to the XY model ground-state energy ($\mid W \mid = \infty$), and the 
ground-state
{\it always} remains in the $Q=0$ sector.
There is no trace of any additionnal transition for
$W >0$, this is in complete agreement with the
DMRG results \cite{SA}.
Let us now consider the $ W < 0$ case.
Below a critical value $W_{c} \simeq -1.75t$ the energy
is quasi linear in W and becomes
very close to the ground state energy of the XY model. 
Hence after the transition
the {\it one particle hopping term is almost frozen out}.
This transition can be observed more clearly in Fig.1b.
Indeed before the transition
 the total momentum of the ground
state is 0 and is $\pi$ after the transition.
This transition from the $Q=0$ to $Q=\pi$ subspace depends only on the parity
of the number of pairs,
therefore the transition is easily observed when this number is odd.

To distinguish superconducting phases corresponding respectively
to $W >0$ and $W <0$, we calculated the distributions of $< n_{k} >$
and $< A^{\dagger}_{Q}A_{Q} >$ in both cases (Fig.2).
For $W >0$ the $< n_{k} >$ distribution continuously goes to
the limiting case $< n_{k} > =1$ (i.e. $| W | =\infty$) and
the distribution of $< A^{\dagger}_{Q}A_{Q} >$ has a strong
peak at $Q=0$.
In the opposite case, for $W < W_{c}$, $< n_{k} > \simeq 1$ (reflecting
the frozening out of one electron hopping term) and 
$< A^{\dagger}_{Q}A_{Q} >$ shows a peak at $Q= \pi$.
For $W >0$, pairs with $Q=0$ appear continuously in the system.
On the other
hand, when $W <0$ the $Q=\pi$ pairs appear spontaneously when
$W <W_{c}$. The competion
between the $t$-term and the $W$-term leads to
the {\it total destruction of the old band structure at the critical point}.

To show that $A^{\dagger}_{Q}$ with $Q=0$ or $\pi$ is a proper operator
to describe the physics of the system we calculate,
\begin{eqnarray}
Z_{Q}= \frac{|<N_{e}-2 | A_{Q} |N_{e} >|}
{(<N_{e}|A^{\dagger}_{Q} A_{Q} |N_{e} >)^{1/2}}
\end{eqnarray}
where $|N_{e} >$ is the ground-state with $N_{e}$ particles.
This overlapp measures the weight of quasi-particle 
with total momentum Q in the ground-state \cite{DS}.
We checked numerically that $Z_{Q}$ is zero for all values
of Q except 0 and $\pi$.
In Fig.3 we calculated this quantity
for $Q=0$ and $Q=\pi$.
For $W >0$, $Z_{\pi}=0$ whilst $Z_{0}$ is finite and continuously
approaches its limiting value.
The jump at $W=0$ is a finite size effect, it can be shown that
it goes as $ 1/\sqrt{L}$.
Thus the usual Cooper pairs appear continuously for $W >0$.

In the opposite case $ W <0 $, $Z_{0}=0$.
Depending on the parity of the number of sites
$Z_{\pi}$ is 0 or 1 for $ W=0^{-}$.

For $L=6$, $Z_{\pi}=0$ before $W= -1.5t$,
the jump which appears at this point corresponds to a 'crossing'
with some excited state, while the final
transition corresponding to
$\eta$-pairing occurs only at $W \simeq -1.75t$, 
after this transition $Z_{\pi} \simeq 1$.

In the case of $L=8$, $Z_{\pi}=1$ for $W=0^{-}$ this is
common when $L=4L_{o}$ due to the presence of
electrons with momentum $p_{F}=\pi/2$.
When we increase $| W |$
the probability to find a  particle on the Fermi surface
is reduced, this is reflected by the reduction of $Z_{\pi}$.
After the transition to $\eta$-pairing $Z_{\pi} \simeq 1$.
This suggests that the $\pi$--pairs appear spontaneously in the ground-state
after the transition.
And $A_{0}$ and $A_{\pi}$ are proper operators
to describe the superconducting states for 
$ W > 0$ and $ W < 0$ respectively.

Another way to visualize
this transition is to show
the pairing phenomenon in real space. The correponding quantity is given by,
\begin{eqnarray}
F(W)=\frac{1}{L} \sum_{i=1}^{L} <n_{i \uparrow}n_{i \downarrow}> -<n_{i \uparrow}>
<n_{i \downarrow}>
\end{eqnarray}
it is plotted in Fig.4a.

This picture shows that for $W >0$ the pairs appear continuously
in the system.
In the opposite case $ W < 0$, and $ | W | \ll t$ the tendancy to pairing
is reduced, this is 
in agreement with the effective repulsive character
of the pair-hopping interaction in this limit \cite{AM}. 
This effective repulsive character remains up
to $| W | < 0.5t$ after which a tendancy to pairing appears.
However the transition to $\eta$-state corresponds
to the jump in $F(W)$ at $W \simeq -1.75t$.
As it is clear from this figure that
the finite size effects have only a {\it weak} influence on this value.

To show that the physics does
not depend on the band filling, we analyse the
model away from half-filling.
In Fig.5 we plotted the lowest energy
as a function of $W$ in two particular cases.
It is clear from this picture that
the transition 
to $\eta$-state remains away from
half-filling.
To analyse the band-filling dependance
of the critical value we calculated
$F(W)$ for a given size $L=10$ and
different number of particles (fig4b).
This picture suggests that the critical
value {\it weakly} depend on the band filling.
When we increase the band filling $W_{c}$
goes from -2t (two particles) to $-1.75t$ (half-filled case).

To conclude, in this paper we have shown that the pair-hopping model
for $W < 0$ 
exhibits a phase transition
into the $\eta$-superconducting state.
The critical value at which the transition takes place
weakly depends on the size of the system and on
the band-filling, it varies between $-2t < W_{c}< -1.75t$.
For $W > 0$ our results are in agreement with
previous works \cite{AM,SA}, and
shows a continuous second-order
transition to usual superconducting state at $W = 0^{+}$, with no
additionnal transition for any $W > 0$.
We argue that this phenomenon will remain 
unchanged in higher
dimensions.
Investigation of the 2D case is
under consideration.

We gratefully acknowledge
W. Brenig, A. Kl\"umper, P. van Dongen,
 D. Vollhardt
and especially A. Kampf and E. M\"uller-Hartmann
for many useful discussions.

Research was performed within the program of the Sonderforschungsbereich 
341 supported by the Deutsche 
Forschungsgemeinschaft. G.J. was partially supported by
INTAS grant N 94-3862.

%
%
\begin{figure}
\caption{Lowest energy vs. W calculated in the $Q=0$ (open circles)
and $Q=\pi$ (open squares) subspaces.
L is the system size.
Full dots corresponds to the XY groundstate.
The transition point is indicated in the figure.
}
\label{f1}
\end{figure}

%
%
\begin{figure}
\caption{$< n_{k} >$ distribution for
different values of the parameter W is plotted in (a).
In (b) $< A^{\dagger}_{k}A_{k} >$ vs k is plotted for
differents values of W (full symbols).
The open symbols corresponds to $W=-\infty$ (open triangles)
and $W=+\infty$ (open squares).
}
\label{f2}
\end{figure}

%
%
\begin{figure}
\caption{$Z_{Q}$ as a function of W for $Q=0$ (a)
and $Q=\pi$ (b).
The transition to $\eta$-pairing state 
occurs at $W_{c} \simeq -1.75t$.
}
\label{f3}
\end{figure}

%
%
\begin{figure}
\caption{F(w) (see the text) as a function of W:
It is calculated at half filling for $L=6,8$ and 10 (a).
The effect of the band-filling for an $L=10$ system
with different particle number $N_{e}$ (b).
}
\label{f4}
\end{figure}
%
%
\begin{figure}
\caption{Lowest energy vs. W calculated in the $Q=0$ (open circles)
and $Q=\pi$ (open squares) subspaces.
L is the system size and $N_{e}$ is the particle number.
}
\label{f5}
\end{figure}

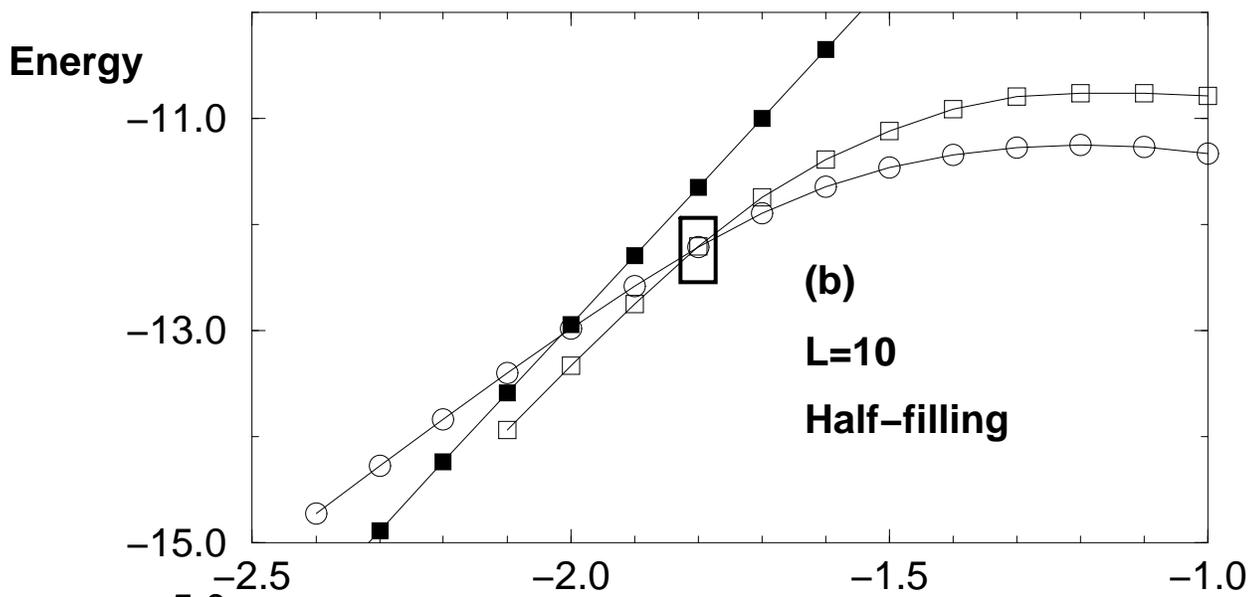
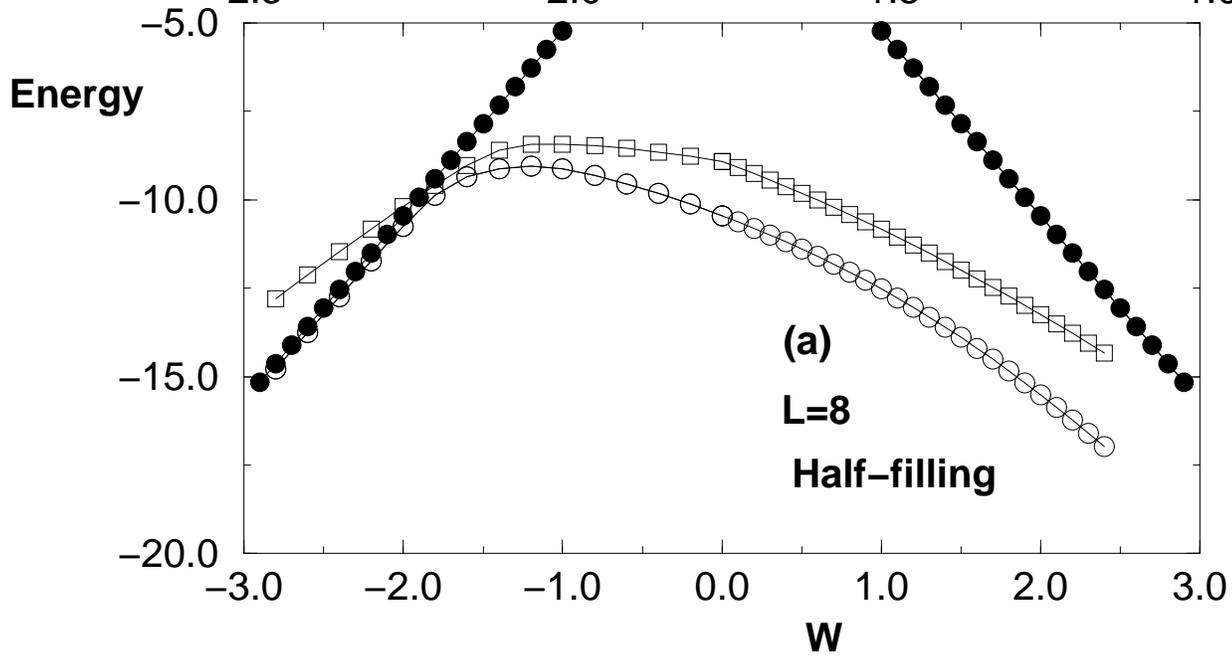
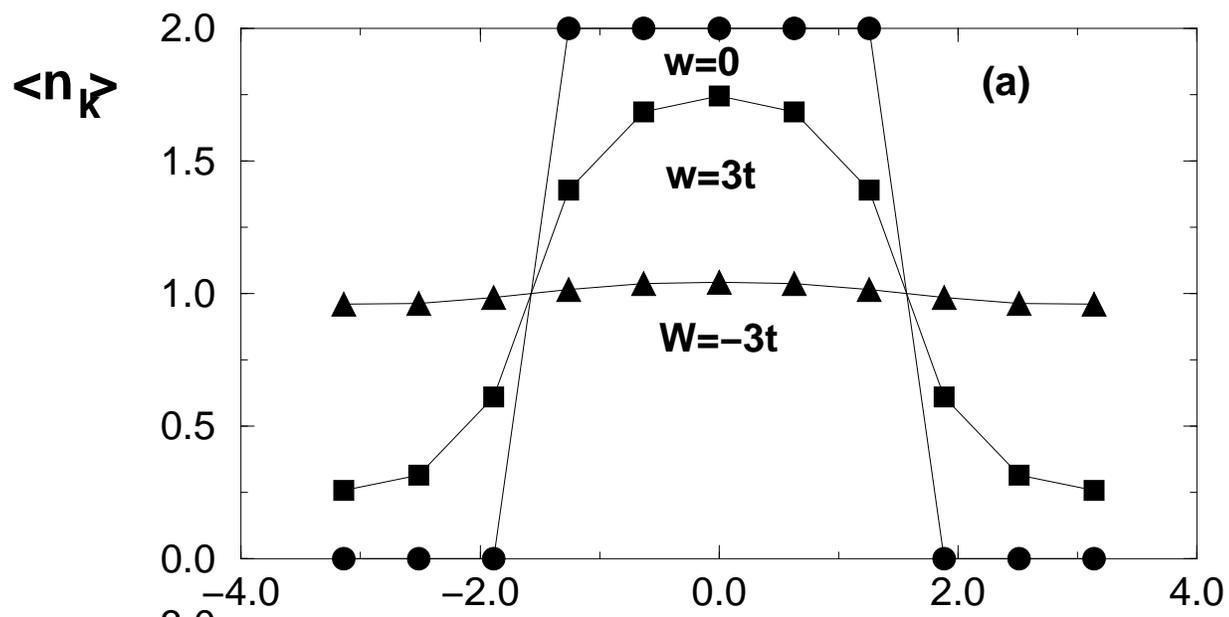
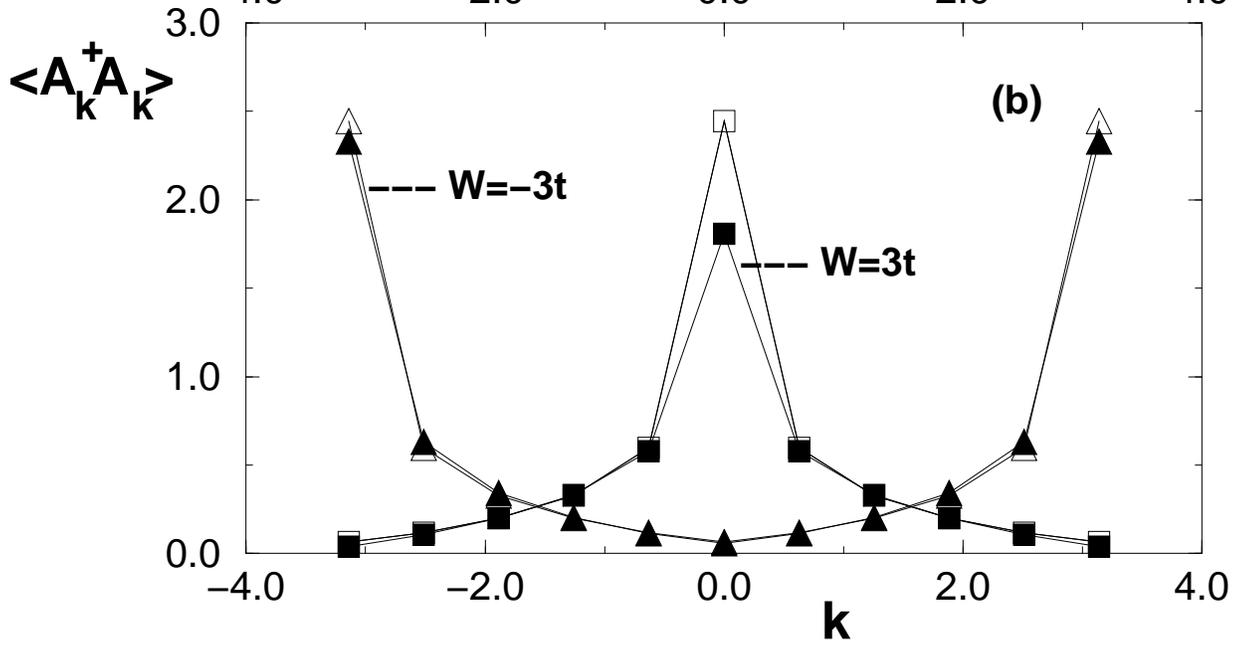
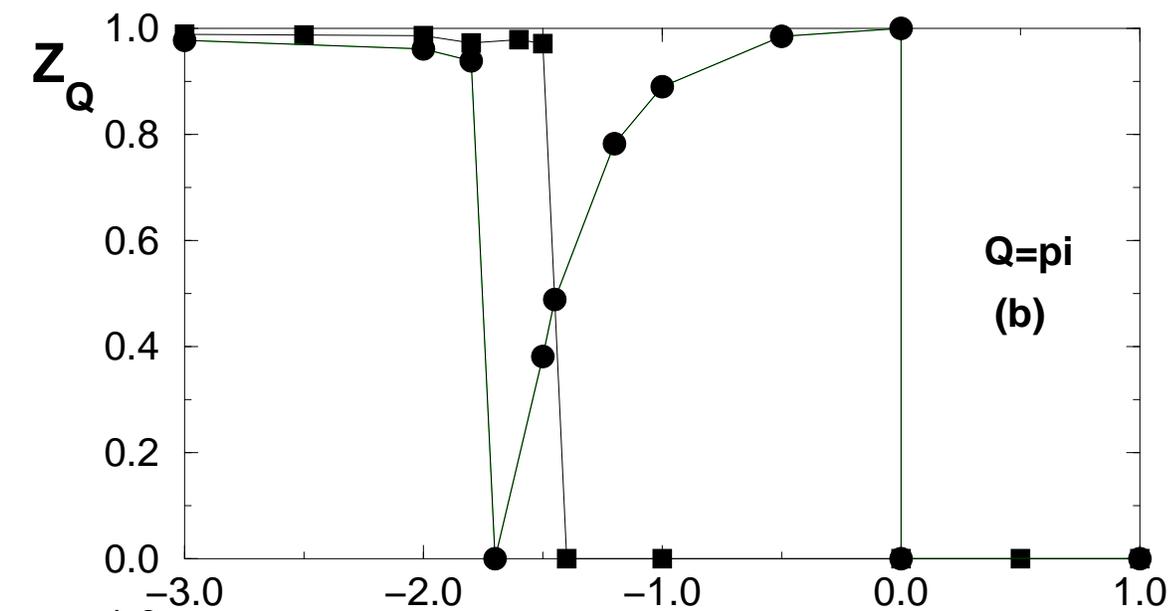
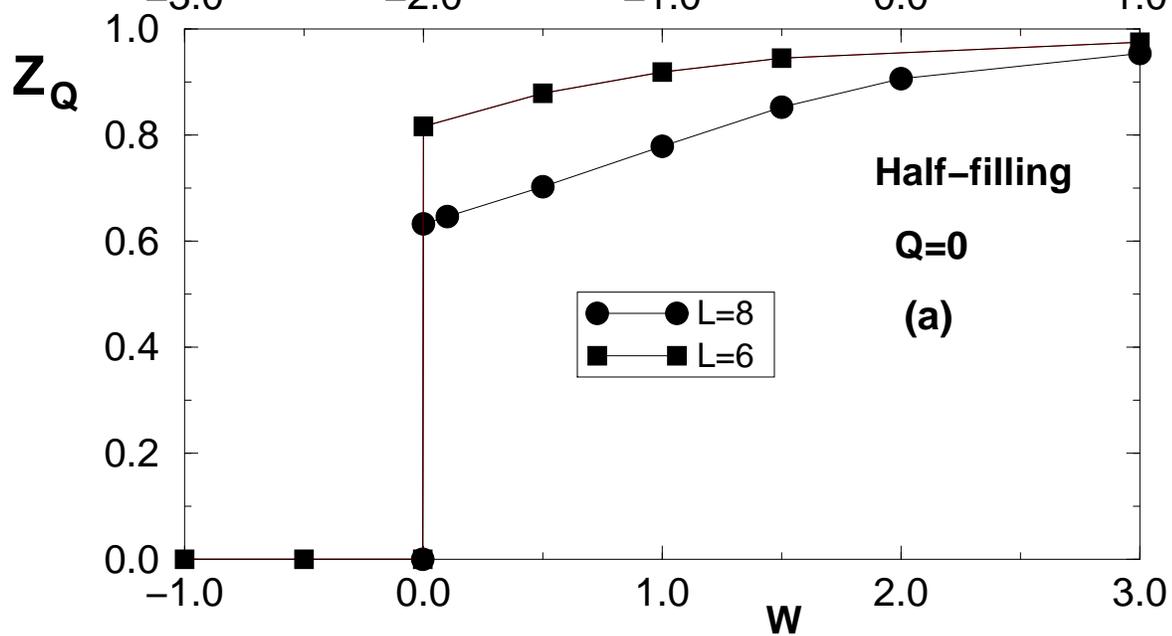
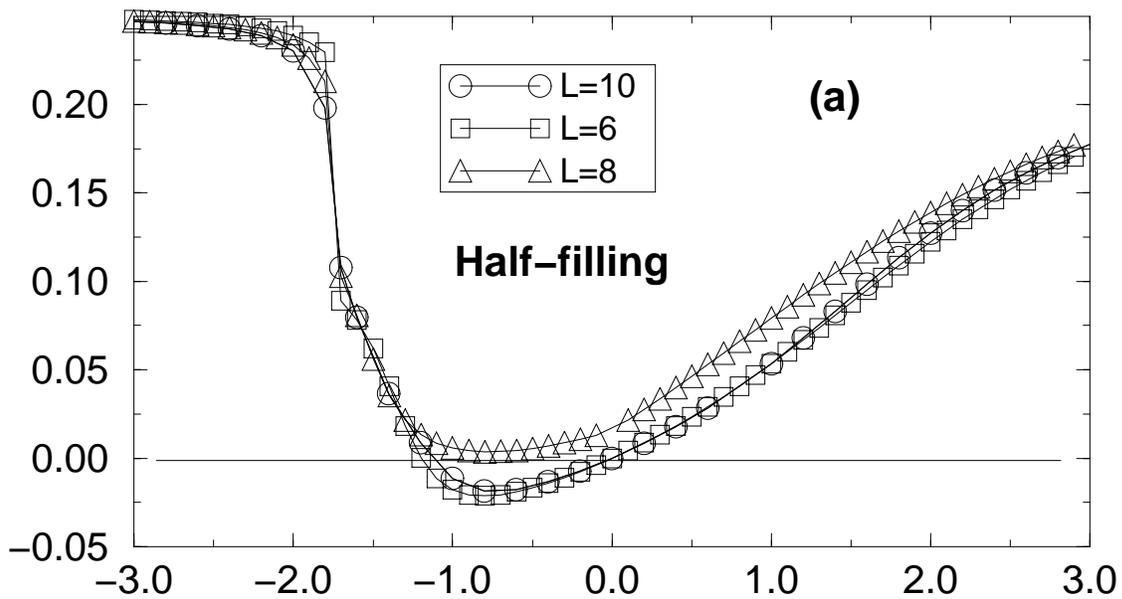
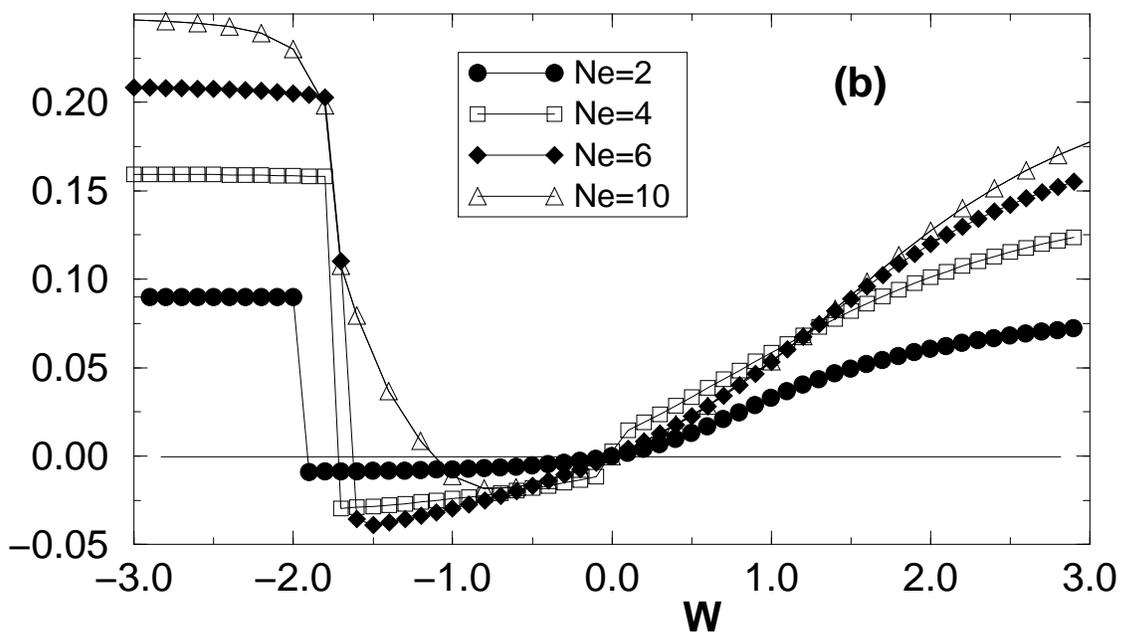
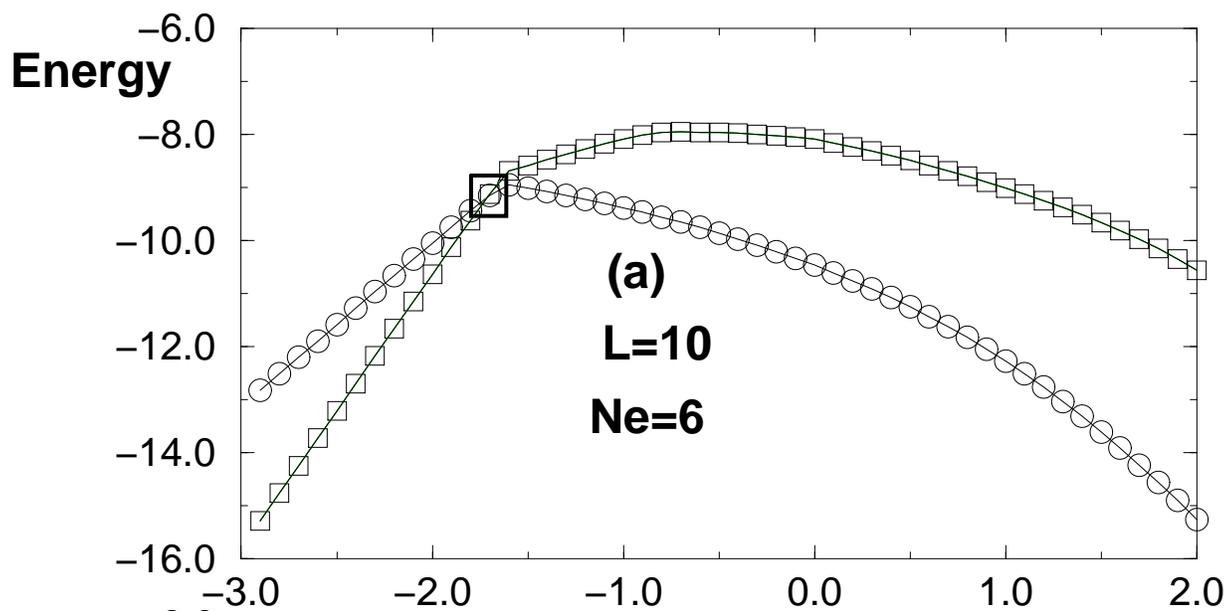
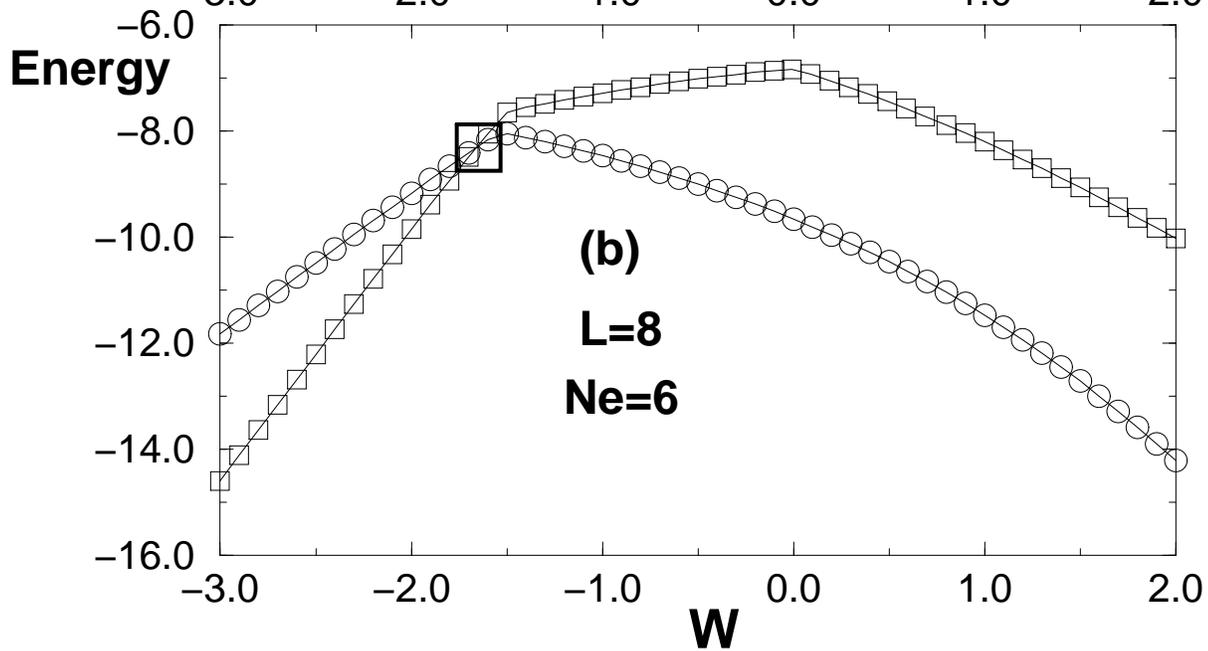
\end{document}